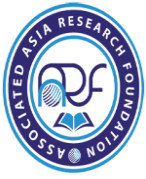

# REVIEW: THE DEVELOPMENT OF NEURAL STEM CELL BIOLOGY AND TECHNOLOGY IN REGENERATIVE MEDICINE

**Divyanjali Shanmuganathan and Nivethika Sivakumaran**

Undergraduate Department of Biomedical Science, International College of Business and Technology, No. 36, De Kretser Place, Bambalapitiya, Sri Lanka

**Abstract**

In the middle of the last century, it has been known that neural stem cells (NSCs) play a key role in regenerative medicine to cure the neurodegenerative disease. This review article covers about the introduction to neural stem cell biology and the isolation, differentiation and transplantation methods/techniques of neural stem cells. The neural stem cells can be transplanted into the human brain in the future to replace the damaged and dead neurons. The highly limited access to embryonic stem cells and ethical issues have escalated the search for other NSC sources. The developing technologies are indicating that it can be achieved before the end of this century. In addition, the differentiation and the maturation of NSCs can artificially accelerate by modern methods.

## 1. INTRODUCTION

Stem cells are the unspecialized cells that have the potential to develop into different kinds of cells found in the body and it acts as an internal repair system. During the division of stem cells, each new cell has the capability either to become another type of cell or remain a stem cell. The remarkable characteristics of stem cells are the ability to renew themselves means making new copies through cell division and it can inactivate after some long period. They regularly divide in the bone marrow to either replace or regenerate damaged tissues also it has the capability to divide inside the Pancreas and Heart under special condition [1].

Totipotent stem cells are the kind of stem cells, which can differentiate into all embryonic cell types and non-embryonic structure cell types such as placenta, and adult cell types, for example, fertilized egg cell- the first cell from this cell is by the cleavage not by stem cells, also they do not self-renew. Pluripotent stem cells can form all the cell types, which can be formed by totipotent except non-embryonic structures like yolk sac and placenta, for example, embryonic stem cells. Multipotent stem cells are found in specialized tissues of fetus and adult. They are capable of forming many tissues but not all the tissue cells of the body. Unipotent cells mean they can only differentiate into one cell type, for example, Spermatogonia can only give rise to sperms [2].

When embryo starts to develop, the totipotent fertilized egg start to moves onto the pluripotent state where the inner cell mass (ICM) of a blastocysts-stage embryo formed and continuously move to specialized cells, multipotent and unipotent cells stage. The Unipotent and multipotent cells are called progenitor cells. The dividing capacity of progenitor cell is limited [2].

According to the regenerative ability of stem cells, the medical industry uses this potential to cure the degenerative neural diseases which are referred as regenerative or reparative medicine [3]. The *in vitro* differentiated human embryonic stem cells (hESCs) is expected to replace the central nervous system (CNS) neurons since this is a most profitable way to suppress and prevent the neuronal disease [4].

Neural stem cells (NSCs) are originating in the adult CNS and produces neural progenitor cells (NPCs) which can give rise to neurons and glial cells known as a combination of oligodendrocytes and astrocytes. To harvest the NSCs scientist have been used few days old embryonic stem cells, umbilical cord blood at the birth and the amniotic fluid. According to ethical issues, it has been banned in many countries to research with fetal tissues so the scientists are urged to find new sources and methods to isolate the NSCs and to differentiate them into neurons *in vitro* to successfully transplant the harvested neurons into a patient [2].

Non-neurogenic regions of the brain cannot produce neural stem cells therefore, the specific region of the brain derived from the donor or the other animal models and micro dissected, cultured with highly enriched culture medium then differentiated and purified *in vitro*. This harvested NSCs used for therapeutic purpose and experiments [5].

In recent years, many different methods have been raised to culture the NSCs and NPCs to differentiate into neurons. I addition, new methods have been developing by researchers to culture and transplant the neurons into patients. There are no exact methods to grafting the laboratory-derived neurons into the recipient. If the neural replacement succeeded in the future, it can cure not only neuro degenerative diseases also psychiatric disorders, spinal cord injury, peripheral injury, blood tumors, epidermal and corneal disorder etc. [3, 6].

## 2. NEURAL STEM CELL BIOLOGY

CNS is the origin of NSCs where two specific regions also are known as "neurogenic niches" give rise to NSCs and they are immature in adult and developing CNS [3]. The major cellular components of adult neurogenic niches are mature neurons, microglia, endothelial cells, ependymal cells, progenitor cells and astrocytes [7].

The two major neurogenic niches in the brain are a Subventricular zone (SVZ) and Hippocampal subgranular zone (SGZ). SGZ located in the dentate gyrus of the hippocampus where NSCs present in the lateral ventricle wall. Exclude these neurogenic niches the other regions of the brain can only generate glial cells [5, 8]. Non-neurogenic regions of the brain are Striatum, Optic nerve, Septum, Spinal cord, and neocortex also proliferating cells can be isolated from these regions and interestingly, they can generate neurons in vitro as well as in vivo when grafted into the neurogenic regions of the recipient [9]. Quiescent is the inactive NSCs (qNSCs) whereas aNSCs are the actively dividing NSCs both live in the neurogenic niches [10].

Stem cells express unique two kinds of genes identified as Sox2 and Oct4. The function of these genes is to promote other stem cell genes and to stop the associated genes of different cell types [11]. NSCs and NPCs are the cells which can proliferate and differentiate into multiple cell types but the proliferative capability is limited and it consists of all undifferentiated progeny of NSCs although it can be unipotent, bipotent or multipotent.

Intriguingly, it did not show the self-renewable ability [12]. According to the expression of unique markers and morphologies, there is two kinds of neural progenitors present in the SGZ which are known as Type 1 hippocampal progenitors (hp) also called as Astro-like NSCs and Type 2 hp. Type 1 hp express the Glial fibrillary acidic protein (GFAP), Astrocyte marker GFAP, transcription factor Sox2 which is important to regulate the stemness. Type 2-hippocampal progenitor is believed that it can be raised from type 1 hp but it is not yet proven. Type 2 hp do not express GFAP and type 2 Sox2 positive cells can self-renew [13].

## 3. NEURAL STEM CELLS AND NEUROGENESIS

### 3.1. Neurogenesis in embryo

Neurogenesis in embryo occurs during embryogenesis. The initial process of neurogenesis is the induction of neuroectoderm later it forms the neural plate. Neural plate folds and makes the neural tube. Radial elongated Neuro epithelial progenitor cells (NEPs) start to appear as heterogeneous population inside the neural structures and they are complex. NEPs undergo symmetric divisions to expand NSC pools in the early stage of neurogenesis. In the second stage, NSCs turn on to asymmetric division cycles along with that generates the lineage-restricted progenitors. Finally, NEPs convert into NSCs and NPCs. Glial restricted progenitors give rise to astrocytes and oligodendrocytes whereas NPCs give rise to neurons [12].

### 3.2. Neurogenesis in adult Mammalia

Neurogenesis in adult referred as a generation of neurons from neural precursors the whole lifetime in restricted brain regions [7]. SGZ born neurons migrate into granule cell layer of the dentate gyrus and convert into dentate granule cells which are the principal neurons and they receive several of inputs from the different areas of the brain via neural peptides and neurotransmitters [13].

SVZ give birth to neurons and they migrate through the rostral migratory stream and develop into granule neurons and periglomerular neurons in the olfactory bulb. In adult Mammalia, the newborn neurons benefit plasticity of the brain and unite with existing circuitry to accept the functions [8, 13].

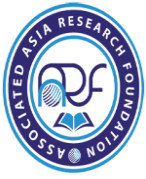

## 4. TRADITIONAL NEURAL STEM CELL CULTURE

**4. 1. Traditional Neural stem cell culture methods**

In the past decades' researchers thought that stem cells can be only harvested from developing embryo because induced pluripotent stem cells (iPSCs) and Embryonic stem cells (ESCs) permit the rise of NSCs population so they have been enhanced several in vitro techniques to differentiate the embryonic stem cell to NSCs [3,14]. The culture conditions for NSCs developed in early 90's by the invention of mitogens; EGF and FGF-2 and this conditions allowed to extend the NSC cell division [3]. In 1981 the first ESCs were isolated from a mouse embryo [15]. Blastocyst stage of the mammalian embryo consists of inner cell mass (ICM) therefore ESCs are isolated from ICM [3].

4.1.1 Mouse embryonic fibroblast feeder layer method (MEF)

James Thomson isolated the first human embryonic stem cells (hESCs) in 1998. This was done by transferring an embryonic ICM which was 3-5 days old onto a MEF layer of the mouse of and he cultured the hESCs on mouse embryonic fibroblast. Later these hESCs maintained in the neurospheres and differentiated into adult NSCs [16].

4.1.2. Technique to generate the neural differentiation in human embryonic stem cell

Nerve growth factor (βNGF) and retinoic acid (RA) can potentially promote the neural differentiation in the ICM of blastocyst stage of the embryo. High concentration of RA and βNGF, dramatically increased the neuronal differentiation and no of neuronal cells meanwhile RA enhanced the neuronal process and production of neurons [14].

4.1.3. Amniotic fluid stem cell (AFS) isolation technique

The rich source of multiple cell types is found in amniotic fluid. The heterogeneous AFS can give rise to multiple cell types. AFS cells can be directed into neural lineage by differentiation with nerve growth factors (NGF). AFS cells were derived from rodent and human AF and isolated by trypsinization method (using trypsin to dissociate cell) and allowed to expand in the culture. Then the suspension poured into NGF containing medium and induced to differentiate into neural cells (i.e. Dopaminergic neuron). Multipotency of AFS cells clones was examined by the retro viral marking [14].

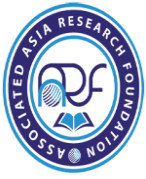

### 4.1.4. Neurosphere culture

Reynolds and Weiss invent this method in 1992 to ensure the presence of NSCs in neurosphere. In addition, it has been using to examine the NSC characteristics such as self-renewal and Multipotency [3]. It is popular for examination of stem cell activity in vitro [17]. The initial step of this assay is a micro dissection of a selected region of CNS derived from Adult CNS, rodent brain or embryo. This is performed by either mechanical or enzymatically procedure. The dissected region is plated in the defined serum-free medium made up of mitogens such as EGF and bFGF. NSCs and NPCs start to divide and form primary neurosphere called "floating aggregates" [3, 12]. After 3 days, it forms small clusters of cells and this clusters detached and re-plated to give rise to the secondary neurospheres. These passages continue for many times to expand the NSC population. Several days after the NSCs and NPCs differentiate into neurons and glial cells then purify and measure the frequency of NSCs [2, 3].

### 4.1.5. Purification Technique-Flow cytometry

This widely using technique is an essential part of NSC culture and purifies NSCs according to its granularity, antigen expression, and size. Some antigen markers can be detectable in this technique, which are EGFR, Nestin, CD 133, Lex, Sox2, and Musashi. This is only beneficial for identification of cell population, which can express markers [17].

### **4.2. Advantages of Traditional Neural stem cell culture methods**

ESCs allow producing vast number of neurons and interneurons *in vitro* and are capable to differentiate into all three germ layers. Compare to adult stem cells, the ESCs are easy to grow easily in culture medium also the population is phenotypically and genetically stable [18]. The benefit of neurosphere assay is testing NSCs for their properties in different conditions also allow to receive and expand aNSCs derivatives to test their properties in variable conditions. Neurosphere can be cultured for many passages up to 60 times and it may vary according to age and origin of tissue species [19].

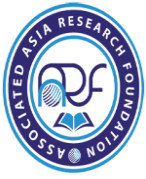

### 4.3. Errors in Traditional Neural stem cell culture methods

In embryonic stem cell culture, the standard protocol for defined culture environment is poorly described [16]. In hESc neural differentiation method, it is difficult to generate homogeneous neuron cell culture due to absent of the lineage selection therefore to avoid this, genetically modified ESCs lines can be used [18]. The hESCs take weeks to form a culture with the daily maintenance so it consumes more time also they grow slowly *in vitro* and unable to decide whether the cells are pluripotent or stable. It is more difficult to assay cell frequency by characterization methods such as gene expression analysis, karyotyping, flow cytometry and immunocytochemistry. This technique provides limited expansion and cells try to lose differentiation and clonal capacity in the long-term passage [14]. Serum or conditioned medium is inappropriate to use in humans so there was a necessary to improve new culture protocols for adult NSCs also these old methods are technically challenging and labor intensive [9,16].

Neurosphere assay also has shown some errors in NSC technique evolution. When aNSCs grow in neurosphere assay, they may get altered in their properties during re-culture for more than 10 times. Unfortunately, quiescent stem cells may not be identified by this assay also it does not allow to examine the bonafide stem cell population. Neurosphere assay has caveat so it cannot be accurately measured NSC frequency because it measures a false frequency of both NSCs and NPCs frequency due to unable of distinguishing the cell types according to this error Neurosphere assay is limited to use as a quantitative in vitro assay [17, 19]. In addition, utilization of hESCs led to the rise of ethical and political controversy. Some countries in the world banned the hESCs associated research. Ethics differ in each country according to their political and religious faith [20]. These ethical issues arise from the technology and the stem cell derivation methods used in the research [2]. The differentiation of pluripotent stem cells from embryo and oocyte is under the ethical policy, therefore, they have been derived from the reprogramming of somatic cells (i.e. Skin cells).These pluripotent stem cells are called induced pluripotent stem cells [20].

The hESCs research is involved in the abortion of human embryos so the application of hESCs is restricted by sharp ethical and political controversy. The high controversy has arisen in the

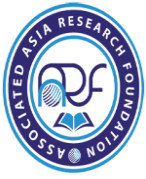

United States about the abortion of embryos. Scientists arguing that the embryo has the potential to become a live-born child when it is implanted into the uterus at the correct hormonal phase. Also argued, when the embryo implanted into a uterus, first it become fetus then only it develop into the live-born child. In some countries, people believe that the embryo has the same moral status as a human individual according to their cultural, religious and spiritual morals, therefore, derivation of embryonic stem cell lines from an embryonic ICM is considered as an extermination [2].

The alternative way is to derive embryos from a couple or women who had the infertility treatment known as In Vitro Fertilization (IVF). After the treatment, there will be some remaining frozen embryos so it can be donated to research rather than decide to destroy them. The couple or women must decide to donate according to the ethical laws and federal regulations. In addition, this is applicable for the fetal tissue donation after the abortion. Pluripotent stem cells can be derived from the donated fetal tissues. Presently, fetal-derived NSCs are used in Batten's disease clinical trials which is a degenerative disease affecting the children [20].

The least ethical restrictions are permitted in several countries such as Belgium, Sweden, and the United Kingdom, to work with pluripotent stem cells but Germany only permit to work with adult stem cells and working with pluripotent stem cell is considered as a crime. Adult stem cells can be used in most of the countries for the research and clinical trials because they do not restrict by ethical controversies [2].

The European Union funded a database to Europe called Human Embryonic Stem Cell Registry (hESCreg) this help to access all the details about hESCs derived project details.

## 5. MODERN NEURAL STEM CELL CULTURE
### 5.1. Modern Neural stem cell culture methods

Ethical issues for hESCs derives methods and lack of technology and stem cell knowledge was led to develop the modern methods [3].

5.1.1. Adherent monolayer culture

The defined serum-free medium used to culture CNS tissues. The medium is boosted with bFGF, EGF, Laminin or fibronectin and Substrate-Poly-L-ornithine, these supplements help in long-term culture of neural precursors. NSCs and NPCs adhere to the medium and forms a monolayer of cells. After the proliferation of cells. An Enzymatic treatment used to detach the cells from the culture surface and replaced another culture medium, which is as same as primary culture, and after long period cells undergo symmetric divisions. Cells induced into differentiating in to neurons and astrocytes by removing mitogens and kept under low serum-containing medium [12].

5.1.2. Neural-colony forming cell assay

It is developed to identify the high proliferative capability of stem cells than progenitor cells [17]. In addition, N-CFCA can distinguish stem cells from the progenitor cells based on their proliferative potential [20]. This is invented by Brent and Louis in 2008 [18]. The procedure starts with the suspension of a single cell into the plate at low cell density in a serum-free medium, which is composed of collagen and other essential factors. The collagen confirms the distinct colonies formed from single cells so this assay also known as the collagen-based semi-solid assay. The medium inhibits the proliferation of NSCs and plated then leave for 21 days to generate colonies and colonies use to read the NSCs frequency. Finally, the equivalent no of colonies derives. The Research concluded that the growth conditions of N-CFCA do not inhibit the proliferation of adult or embryonic precursors. Instead of that, the time 21days enabled the high proliferative potential of cells to generate colonies with wide size distribution [21].

**5.2. Errors in Modern Neural stem cell culture methods**

Adherent monolayer culture is not well organized to maintain NSCs *in vitro* and it produces the heterogeneous population of a cell, which is inappropriate for the transplantation [12]. N-CFCA takes a long time to form colonies; therefore, it consumes time, labor cost and needs daily attention to maintain [21].

## 6. ADVANCED NEURAL STEM CELL CULTURE METHODS

Ongoing researchers and recent inventions play a major role in neural stem cell technology.

### 6.1. Reprogramming of skin cells into induced pluripotent stem cells

Very recently, the scientist used a technique to produce stem cells by reprogramming the skin cells to screen for tau-targeted therapy. The produced stem cells could differentiate into any kind of cell. The iPSCs are the most promising way to regenerate the Alzheimer's disease and to produce a homogeneous population of neurons [11].

### 6.2. Hydrogel technique to promote neurogenesis

In 2071, hydrogel used to induce the production of neurons. This method helps to differentiate NSCs into neurons and induce the rate of maturation. Hydrogel facilitates in its consistency and structure to modify and create similar properties of the human brain and this biomaterial has some elasticity as cerebral tissue. In addition, it contains the synthetic adhesive molecules "IKVAV" to accelerate the process of neuronal fate and neurogenesis and increases the chances to give rise to new neurons from NSCs. Hydrogel stiffness is as same as human brain so it facilitates the stimulation of neural development. The stiffness of hydrogel differently affects neurogenesis of embryonic and adult NSCs. The NPCs of embryo and adult stem cell progenitors accelerated by adjusting the hydrogel stiffness. The Bifunctional hydrogel is produced by combining the laminin peptide IKVAV renders and polylysine [22].

### 6.3. Clustered Regularly Interspaced Short Palindromic Repeats-genome activation method

It is created to produce a large number of human brain cells to ease drug discovery. Scientists of Gladstone institute convert skin cells into stem cells by activating a specific gene using CRISPR technology and they induced the pluripotent cells from skin cells with the help of four key proteins those are transcription factors. They alter the responsible gene expression. The transcription factors switch off the skin cell-associated genes and switch on the genes that are responsible for stem cells. Thousands of genomic locations can be selected by one transcription factor and alter the gene expression at each location. Sox2 and Oct 4 are the responsible genes to switch on other stem genes and to switch off those genes which are associated with other cell types [11].

This technique is an effective tool to create definite manipulation of a genome by the selection of unique DNA sequence. This single location target on the genome induce the natural chain reaction and result in reprogramming of skin cell into iPSCs [11].

### 6.4. Developing neural stem cell delivery method

The delivery of harvested stem cells into therapy is limited and undeveloped. The limitation is due to the inaccurate injection site, the inability of cells to survive, inability to skip puncturing structures such as blood vessels, which create hemorrhage when penetrating into the brain. Scientist developing MRI-guided NSC delivering method to transplant the NSCs and it is expecting to solve the problems. This method is still at the research level and researchers working hard to make it successive [20].

## 7. ROLE OF NEURAL STEM CELLS IN NEURODEGENERATIVE DISEASES

The common neurodegenerative disease is Parkinson's disease (PD), Prion disease, Alzheimer's disease (AD), motor neuron disease (MND), Spin cerebellar ataxia, Spinal muscular atrophy and Huntington's disease.

### 7.1. Parkinson's disease

PD affects the area of brain named as 'substansia nigra 'where dopaminergic neurons are rising. The dopamine is secreted by dopaminergic neurons which is important in contributing to a control movement pattern. PD patients may produce less dopamine or no dopamine due to degeneration of dopaminergic neuron consequently the dopaminergic neurons start to die and patients may experience the symptoms such as tremor, bradykinesia (slowed movement), and speech difficulty, write difficulty , rigid muscle, and constipation [23].

Scientists hope that the human dopaminergic neurons derived from the culture can be transplanted into PD patient. The transplanted neurons can be replaced the dopamine level and relieve symptoms of the disease also it will break the traditional treatments such as deep brain stimulation and gene therapy. This idea was researched in 1994 by isolating fetal ventral mesencephalic tissue from aborted fetus which consists of high no of dopaminergic neuroblasts and was transplanted into PD patient. The grafted dopaminergic neurons survived and re-innervated the striatum, unfortunately, the patient was led to the induced dyskinesia and this could be due to dopamine read over sensitized receptor or due to the non-dopaminergic region of the graft. According to ethical issues fetal-derived methods cannot be examined to solve this problem [23].

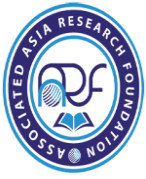

In 1995, Lorenz Studer cultured the dopaminergic cells from dividing progenitors from the fetus and it was transplanted into rats. After some days, symptoms were improved in the rat, which is affected by PD.Lorenz tried to transplant the cells into monkeys but he couldn't send neurons from their birth region to effective region (the long extensions). He could be able to do this in rodent brain but that wasn't easy in the animal brain because the brain is more complex and has long extensions. Nowadays researchers trying to transplant the neurons at their site of action instead transplanting into substansia nigra also they are working on to graft the NSCs into patients by performing advanced research techniques [2].

### 7.2. Alzheimer's disease

Alzheimer's disease is a neuropsychiatric disorder also a neurodegenerative form of dementia (set of symptoms of memory loss and ability in thinking) and it caused by neurofibrillary tangles (NFTs) which led to the neuroinflammation and neuronal loss. The symptoms of this disease are confusion, loss of learning skill, difficulty in thinking and writing and increased loss of memory [20].

NSCs can be used to replace the lost neurons to regenerate. This may rebuild the lost key proteins and the neurons. The challenges of this transplantation therapy are epigenetic and genetic backgrounds of donor cells. The epigenetic memory of donor cell may cause abnormal cell stability gene expression. The Genetic defects can be caused some biochemical changes in the AD so before, glial and neuronal transplantation the DNA editing is necessary. CRISPR molecular scissor could be used to edit the donor's cell [11, 24].

It is difficult to select the specific location for transplantation throughout the CNS in AD patient because the false introduction may lead to tumors and rejection risks [20].

### 7.3. Motor neuron disease

The selective death of motor neurons occurs in this disease due to altered glial function, mitochondrial dysfunction, oxidative stress and other mechanisms. Unfortunately, non-neural cells also contributing to this disease mechanism [25].

Motor neurons can be isolated from human and rodent fetus this is a golden tool to validate the research but due to ethical issues, it is prohibited. Recently, mouse ESCs-derived motor neurons transplanted into the chick embryo and surprisingly, a tiny amount of axons extended into the periphery and formed neuromuscular junctions [25].

## 8. CONCLUSIONS

NSCs have the potential to give rise to the neurons and glial cells so it was studied to generate in vitro and cultured to replace into a human brain. It is thought that the replaced neurons can work normally inside the brain and it will replace the degenerated neurons, unfortunately, there are no exact safety methods to transplant the cultured cells into the brain. The complexity and the long extensions of the mammalian brain give more challenges to be succeeded in this field.

The restriction in fetal-derived stem cells for research delays and limit the development of new successive techniques to cure the disease, even though there are some least restricted ongoing research performed with human embryonic stem cells, which are legally derived from the donors. Recently scientists could produce homogeneous neuronal cells from genetically modified hESCs. One of the breathtaking successful is the derivation of stem cells from the non-neurogenic regions of the brain. Although several different sources have been used to isolate the NSCs among them the reprogrammed skin cell is more effective than the other sources because it is quicker, cheaper and can give more effective iPSCs. The protocols to expand the NSCs in vitro have been developing while the culture medium to grow and differentiate NSCs been improving and developing every year.

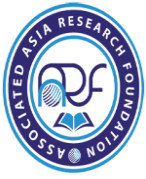

**International Research Journal of Natural and Applied Sciences**
ISSN: (2349-4077)
**Impact Factor**- 5.46, Volume 5, Issue 3, March 2018
**Website-** www.aarf.asia, Email : editor@aarf.asia ,
editoraarf@gmail.comThe Scientist working on a promising research to develop MRI-guided stem cell delivery method but before using on therapeutic purpose, scientists should unsolved the technical and neuropharmacological problems. They should develop the most appropriate in vitro methods to manipulate, to obtain the NSCs, and to derive the homogeneous culture of neuronal cells. They should investigate the no of cells, which are needed to be transplanted on the specific disease stage. Researchers should investigate the pharmacodynamics and pharmacokinetic effects of a mouse after the transplantation of neuronal cells into a diseased mouse.

The field is moving forward to its destiny, new trials are continuously being planned and performed but none was successful. The neural stem cell therapy may get valid in the clinical therapeutics in the next decade.

**ACKNOWLEDGMENT**

The authors wish to thank the International College of Business and Technology, Sri Lanka for the financial support.**REFERENCES**

[1] Various, 2006. The stem cells collection, 1st ed, Science. AAS Office of publishing, Bedford.

[2] Mummery, C., Van de Stolpe, A., Roelen, B. and Clevers, H.2014. Stem Cells. 2nd ed. Burlington: Elsevier Science.

[3] Casarosa, S., Bozzi, Y., Conti, L., 2014. Neural stem cells: ready for therapeutic applications? Mol. Cell. Ther. 2, 31. https://doi.org/10.1186/2052-8426-2-31

[4] Schuldiner, M., Eiges, R., Eden, a, Yanuka, O., Itskovitz-Eldor, J., Goldstein, R.S., Benvenisty, N., 2001. Induced neuronal differentiation of human embryonic stem cells. Brain Res. 913, 201–205. https://doi.org/10.1016/S0006-8993(01)02776-7

[5] Song, S.H., Stevens, C.F., Gage, F.H., 2002. Astroglia induce neurogenesis from adult14

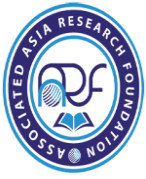